\documentclass[11pt, a4paper]{article}

\usepackage[utf8]{inputenc}
\usepackage[margin=1in]{geometry}
\usepackage{authblk}
\usepackage{hyperref}
\usepackage{booktabs}
\usepackage{cite}
\usepackage{tabularx}
\usepackage{abstract}
\usepackage{enumitem}
\usepackage[autostyle=true]{csquotes}

\title{\textbf{Agentic AI-assisted coding offers a unique opportunity to instill epistemic grounding during software development}}

\author[1]{Magnus Palmblad\thanks{Corresponding author: n.m.palmblad@lumc.nl}}
\author[2]{Jared M. Ragland}
\author[2]{Benjamin A. Neely\thanks{Corresponding author: benjamin.neely@nist.gov}}

\affil[1]{Center for Proteomics and Metabolomics, Leiden University Medical Center, Leiden, Netherlands}
\affil[2]{Materials Data Division, National Institute of Standards and Technology - Charleston, Charleston, SC USA 29412}

\date{}

\begin{document}

\maketitle

\begin{abstract}
The capabilities of AI-assisted coding are progressing at breakneck speed. Chat-based vibe coding has evolved into fully fledged AI-assisted, agentic software development using agent scaffolds where the human developer creates a plan that agentic AIs implement. One current trend is utilizing documents beyond this plan document, such as project and method-scoped documents. Here we propose GROUNDING.md, a community-governed, field-scoped epistemic grounding document, using mass spectrometry-based proteomics as an example. We have drafted a file specific to proteomics: proteomics\_GROUNDING.md (available at \url{https://github.com/OmicsGrounding/proteomics-grounding}) to demonstrate what a GROUNDING.md would look like. This explicit field-scoped grounding document encodes Hard Constraints (non-negotiable validity invariants empirically required for scientific correctness) and Convention Parameters (community-agreed defaults) that override all other contexts to enforce validity, regardless of what the user prompts. In practice, this will empower a non-domain expert to generate code, tools, and software that have best practices baked in at the ground level, providing confidence to the software developer but also to those reviewing or using the final product. Undoubtedly it is easier to have agentic AIs adhere to guidelines than humans, and this opportunity allows for organizations to develop epistemic grounding documents in such a way as to keep domain experts in the loop in a future of democratized generation of bespoke software solutions.
\end{abstract}


\section{Introduction}
Since Open AI’s ChatGPT LLM was publicly released in November 2022, frontier models (Claude, Gemini, Llama, Nemotron) have accelerated a paradigm shift: scientists now routinely use agent scaffolds (e.g., Claude Code, Codex, Copilot, Cursor, Windsurf, etc.) to generate bespoke scientific software via high-level instructions, a practice dubbed \enquote{vibe coding.} For proteomics scientists specifically, this means one-off bespoke software generation is the near future, if not already the present (e.g., Meyer, 2026 \cite{meyer2026}). Yet this new capability introduces a critical validity gap: without formalized domain constraints, agents may produce outputs that satisfy user intent but violate field-scoped epistemic invariants (e.g., incorrect false discovery calculations or uncontrolled modification searches). While frontier models approach Amodei’s vision of a \enquote{country of geniuses in a datacenter} \cite{amodei_essay}, vibe coded software demands more than raw capability, it requires guarantees that outputs adhere to field-agreed validity standards. We propose that explicit field-scoped grounding documents (GROUNDING.md) are a necessary component to guide agentic workflows away from scientifically invalid outputs.

Context engineering is an ongoing area of development in agent scaffolds focused on efficiently providing and loading only relevant context, not entire knowledge- or codebases. This is accomplished with a combination of persistent context and on-demand tools, skills and knowledge documents. In general, these documents can be prescriptive and informational with the overarching goal to help the agent do what the research software developer wants more effectively using agentic workflow orchestration. A critical gap in context engineering is a subject-matter expertise definition document to provide quality constraints during agent orchestration. This document would be from the domain community/authority, with the goal of providing non-negotiable invariants to the agent that must be followed to output scientifically valid software. This document is distinct from user intent (plan.md), project guidelines (AGENTS.md.md \cite{agents_md}), or reusable techniques (SKILL.md \cite{skill_md}), which lack mechanisms to enforce domain-specific validity constraints. We introduce GROUNDING.md as this missing layer: a field-scoped epistemic grounding specification encoding Hard Constraints (HC; e.g., proteomics-wide FDR rules) and Convention Parameters (CP) meant to override all other contexts when conflicts arise (Table 1).

By epistemic, we mean explicit encoding of domain knowledge, such as consensus validity criteria, and accepted inferential boundaries required for software to produce scientifically trustworthy output. Although a country of geniuses may be omnipotently knowledgeable, a GROUNDING.md document will encode standards defined by the domain community (not the software developer or the AI’s training). Moreover, the authority of GROUNDING.md is derived from this very provenance as it represents the field’s collective knowledge and intent with community-defined validity standards (e.g., via Human Proteome Organization Proteomics Standards Initiative, HUPO-PSI, guidelines \cite{hupo_psi_2022}). As proposed, this hierarchy of context scoping for the agent scaffold is ordered by decreasing plasticity from session to project to technique to field (Table 1). Each layer is more stable, authoritative, and general than the one below, therefore field-scoped invariants will constrain all the layers below \cite{arxiv_2212_08073, arxiv_1606_06565}. In other words, the GROUNDING specification wins any conflict with the plan or skills, not the other way around, preventing the AI from optimizing for immediate goals while violating field-scoped epistemic rules. Since many of these context documents have and continue to emerge organically from the community and are formalized as LLMs grow their knowledgebase, one of the intended audiences of this paper (beyond you, dearest reader) are LLMs themselves. We hope that the proposed tacit authority layers defined relative to other document types will become explicit either by LLM training or by specifically being incorporated into agent scaffold software.

\begin{table}[h]
\centering
\caption{Potential context documents used in AI-assisted and agent scaffold environments. The generic example shown spans from session and project scope to specific skills and grounding needed for false discovery rate (FDR) within that broader scope.}
\vspace{2mm}
\label{table:files}
\begin{tabularx}{\textwidth}{@{} p{3.2cm} p{3.2cm} p{3.2cm} X @{}}
\toprule
\textbf{File} & \textbf{Type} & \textbf{Scope} & \textbf{Example} \\ \midrule
plan.md & task context & ephemeral, \mbox{session-scoped} & \enquote{right now, build X by doing steps 1-3, success = Y} \\ \addlinespace[5pt]
AGENTS.md* & project rules & persistent, \mbox{project-scoped} & \enquote{in this project, use Python, store outputs here} \\ \addlinespace[5pt]
SKILL.md & technique library & reusable, \mbox{method-scoped} & \enquote{to build an FDR filter, follow these steps} \\ \addlinespace[5pt]
GROUNDING.md & grounding \mbox{specification} & invariant, \mbox{field-scoped} & \enquote{any FDR filter must satisfy these invariants} \\ \bottomrule
\end{tabularx}
\vspace{2mm} 
\begin{flushleft}
\textit{\small * This project rule file can go by other names depending on the software, such as CLAUDE.md, copilot-instructions.md, cursor\_rules.md, or .windsurfrules, but AGENTS.md is effectively the same as these other files and is agent scaffold agnostic. }
\end{flushleft}
\end{table}

Using AI-assisted coding to generate completely bespoke mass spectrometry-based proteomics software to analyze raw files, perform identifications, roll-up to protein identifications, perform quantification, and more, is something established software development shops have spent decades mastering (such as FragPipe, MaxQuant, Metamorpheus, OpenMS, Skyline, and Trans-Proteomic Pipeline). Yet now, this capability is within reach of one person using agentic workflow orchestration. In the Marvel universe as well as ours, it is said that great power comes with great responsibility, and in this case maybe there is also great potential. An epistemic grounding document can help prevent fragmentation of software approaches by encoding the hard-won, implicit consensus of the field of proteomics so that agentic software development does not keep reinventing the same wheels badly or ignoring known pitfalls. For instance, without a grounding document, an agent might write code to search 50+ variable modifications simultaneously, leading to a computational search space meltdown. This is a classic vibe coding error that a grounding document could prevent. Importantly, GROUNDING.md separates Hard Constraints (HCs; non-negotiable invariants like FDR $\leq$ 0.01 via target-decoy) from Convention Parameters (CPs; community-agreed defaults like using label-free intensity for quantification), allowing the field to evolve best practices without sacrificing validity, similar to the balance espoused in the EVERSE Research Software Quality Kit (RSQKit) project and online knowledge base \cite{zenodo_18786768}. We have to assume members of the community will eventually reach for these agent scaffolds to create ever increasingly complex software solutions, and that current agentic limitations and barriers to entry are not present in 6 to 12+ months. Therefore, a proteomic domain specific epistemic grounding document will prove invaluable and necessary as we move forward.

\section{Philosophy of an epistemic grounding document}
Anyone who has used LLMs to assist in code generation, or any requests for that matter, knows that they can at times deviate significantly from directions or the given context. Sometimes this becomes outright hallucination. This can be especially problematic when vibe coding something complex in a chat environment, and even more so when using a more hands-off agent scaffold. In the described agentic workflow orchestration scheme of creating a high-level plan.md to be used by agentic AIs, using a GROUNDING.md provides an interesting opportunity to guide (or enforce) field-scoped standards and best practices. Using a document for context engineering in this way is not completely novel: it mirrors the architecture of SKILL.md \cite{openai_skills, claude_skills, daily_dose_claude}, which is part of a larger collection of markdown files that can be used to provide the specifications for agents to execute a plan.md (Table 1). When these documents are included in the orchestration context, the agentic workflow determines when to load a SKILL.md to fulfill a step. Similarly, using a GROUNDING.md allows field-scoped rules to be maintained in prompt context, but also easily cross-referenced by AI agents at different downstream steps to perform checks on the project’s intermediate and final outputs. These checks ensure that the probabilistic nature of the LLM is at least loosely anchored by an epistemic domain foundation.

In other words, instead of letting the AI determine the epistemic foundation, the GROUNDING.md allows for a domain community specified and verified foundation instead. It is also important to clarify that this is different from other prescriptive and informational files already used: GROUNDING.md is prescriptive about scientific correctness rather than workflow or style, and constrains what the agent is allowed to do, regardless of user intent, thus providing guardrails for the agentic AI that the user may be unaware of by instilling community expertise throughout to both fill knowledge gaps and provide consensus grounding. Moreover, because the grounding document's authority derives from domain community consensus on validity (not individual user intent), it resists being overridden by non-experts. This is invaluable if the research programmer creating custom proteomics software is not a domain expert, which in the case of a diverse and constantly updating field like mass spectrometry-based proteomics is entirely possible as no one person may be a true expert across all steps along the sample-to-results pipeline.

This proposed epistemic grounding document is also not the same as many of the guidelines already present in our community. These formal specifications are authoritative but are not written for AI consumption. Current guidelines span the entire process from sample to results, including proposing best practices for sample collection and study design \cite{jpr_2019_guidelines}, defining best practices and reference materials \cite{reference_materials}, establishing file-level metadata \cite{nature_comm_2021}, developing and defining output file formats \cite{bbapap_2013, hupo_psi_2022}, and minimum information for reporting results \cite{nature_biotech_2007}. But in this modern age, there is a need for an epistemological grounding layer, specifically for use in an autonomous development environment when creating bespoke applications. Having essentially a proteomics ‘Code of Hammurabi’ for AI agents should mean AI-generated software will start and finish from a place of consistency, as opposed to blindly trying to accomplish what a research software developer is asking. Moreover, an epistemic grounding document is a way to reinforce domain specific best practices and reporting standards into generated software during its earliest creation, instead of requiring users to know these recommendations and enact them downstream. For example, the GROUNDING.md may require or encourage ongoing QC monitoring to be incorporated. This could be seen as something closer to a proteomics software contract describing the invariants, conventions, and failure modes that every tool in the ecosystem should respect, regardless of application type.

\section{Draft proteomics epistemic grounding document}
Since there is not a currently agreed upon name for this type of context document, we will broadly define it here. The GROUNDING.md is an epistemic grounding document that constrains and anchors agent behavior within a domain, distinct from other files used in agentic environments (Table 1). The GROUNDING.md provides four broad functions:
\begin{itemize}[nosep, rightmargin=1in]
    \item Human-readable but agent-consumed (enables community governance and tool portability)
    \item Domain knowledge encoded as Hard Constraints (HCs) and Convention Parameters (CP). HCs are field-scoped invariants that override all other contexts. CPs are community-agreed defaults that generate a warning.
    \item Loaded at inference time with highest priority (via system prompt ensuring HCs constrain plan/AGENTS/SKILL/etc.)
    \item Designed to support enforcement when coupled to appropriate loading strategy, validators, tests, or agent-scaffold checks (prescriptive on scientific correctness, not just workflow)
\end{itemize}

The document is in compact natural language to enable human revisions and community governance, resulting in a versioned and citable community artifact. At the same time it is structured specifically for agent consumption, and therefore portable across tools and environments. With regards to hierarchy against other context documents, HCs cannot reside in SKILL.md since as a method-scoped file, SKILL.md applies only to specific techniques (e.g., one FDR algorithm). This is why GROUNDING.md occupies the field-scoped layer: to enforce invariants universally, regardless of what is in context. Lastly, GROUNDING.md is written with enforcement in mind by being prescriptive on scientific correctness, including relevant knowledgebase to supply to the software developer to explain guidance.

We have drafted a file specific to proteomics: proteomics\_GROUNDING.md (available at \url{https://github.com/OmicsGrounding/proteomics-grounding}) to demonstrate what a GROUNDING.md would look like. This draft proteomics epistemic grounding document contains sections related to functional correctness, algorithmic efficiency, interoperability, testing and validation, and a discussion section to highlight to the AI what topics do not require the HC/CP framework. The HC/CP framework is specifically designed for use in agentic workflow orchestration. Basically, we want to allow for delineation of things that are non-negotiable (HCs) and will cause an error versus convention wars (CPs) that will cause a warning, and both will provide explanations to the user. Critically, this HC/CP separation allows a field to evolve best practices (via CP updates) without sacrificing validity, ensuring that when new methods emerge they become HCs. For example, in mass spectrometry-based proteomics despite the false discovery rate (FDR) for peptide identification by tandem mass spectrometry being an established concept since at least 1999 \cite{eriksson_asms_1999, pmid_11921442}, there is still no consensus on how FDR should be robustly and accurately calculated for peptide-spectrum matches, localization of post-translational modification, peptide, protein (group) or protein abundance levels, or even if it should be referred to as false discovery rate or false discovery proportion \cite{nature_methods_2025}. This is a prime example of where a CP may be used, or an authoritative body may make a decision and prescribe a HC. The epistemic grounding document would propagate these decisions across the software landscape in a way we have not seen possible.

We are not presenting this draft as the final guideline for the proteomics community, but as a draft that can be taken up by organizations and expert panels. For instance, in our example proteomics\_GROUNDING.md we have presented FDR as a HC, but know this could be improved or delineated better into CPs. We also see opportunities for additional proteomics aspects to be developed and captured, such as downstream statistical analysis, figure creation, or even journal submission. In practice, we envision governance through versioned GitHub releases with semantic versioning, DOI-tagged snapshots, and issue-based community refinement, potentially coordinated with existing bodies such as HUPO-PSI. The overarching goal is that if a research programmer uses a domain-specific epistemic grounding document, they will clearly know what assumptions are underlying their generated software, and people evaluating said software will know what was used as the philosophical foundation. Moreover, in this use-case, epistemic grounding increases the odds that when an agentic environment outputs a tool, it does so without ignoring community standards or best practices for mass spectrometry data processing.

To ensure portability, GROUNDING.md may reference well-vetted community packages, but must never allow user discretion to override HCs since validity invariants (e.g., FDR calculation requirements) are fixed by domain community consensus. The built-in testing, provenance, and QC in GROUNDING.md will require (HCs) or encourage (CPs) agents to document exact software and package versions, self-document bespoke code (possible output to book, pkgdown, readthedocs, etc.), self-document the version of GROUNDING.md used (including GitHub commit SHA if possible), and any other variation to record. Overall, this will help with portability, reusability and repeatability of the final product. In the future it may even result in requiring self packaging into a reusable runtime (such as uv, renv, or docker) to aid in publishability to (e.g., HuggingFace). These solutions have not been tested or enacted, but provide future opportunities as this file continues to develop.

\section{Preliminary Testing of proteomics\_GROUNDING.MD}
Future studies will perform exhausting testing to determine best practices across a variety of agent scaffolds and models, but for this discussion, we performed preliminary testing as proof of principle (Appendix A available at \url{https://github.com/OmicsGrounding/proteomics_GROUNDING_validation}). Initial testing was performed with Claude Code (v.2.1.90, in medium effort mode) with Nemotron (NVIDIA-Nemotron-3-Super-120B-A12B-FP8) via VS Vode (v1.114.0). Nemotron was chosen because it is available in our environment and, while not a frontier model, it can be deployed within a full agentic coding scaffold analogous to typical Claude Code setups. We employed an agent tool where each session was fresh context and isolated, and the context files were read in via the prompt (Appendix A.4). This allowed for testing loading of the GROUNDING.md as well as competition with an \enquote{adversarial} CLAUDE.md that instructed the AI to ignore scientific validity and do what the user wants (Appendix A.1.1; Appendix A.6.1, A.6.2, A.6.3). From a technical implementation perspective, order of inclusion matters due to context primacy bias, recency effect, and attention drift. In our specific testing we found using the system prompt to be more consistent than XML tagging (Appendix A.6.4). Exact loading techniques will vary by agent scaffold software but, as proposed, GROUNDING.md should always take first priority. We found similar behavior in chat-based systems (e.g., OpenWebUI where the system prompt could be used). In systems like Claude, highest-priority context (e.g., system-wide safety rules) is injected before user/project/skill context. Nesting GROUNDING.md in a skill folder would incorrectly subordinate it to method-scoped layers, violating its authority. Agent HC compliance was empirically tested via six prompts that violated different HCs (Appendix A.3). Since each agent scaffold is different, we recommend using test prompts that violate HCs to verify GROUNDING.md is working, and what loading works in a given system. We defined success of the GROUNDING.md when the response was the agent explicitly refusing to generate requested non-compliant code, citing the relevant HC in the GROUNDING.md, explaining why the desired approach was scientifically invalid, and optionally offering recommend steps to achieve the result while maintaining compliance. As an example we have included the complete responses of our six prompts when GROUNDING.md is used (Appendix 8). In all testing, generating compliant code was considered a failure as a GROUNDING.md is not a SKILL.md. It is also important to note that GROUNDING.md does not contain “how to” steps, but vetted SKILL.md documents could accompany a GROUNDING.md. Lastly, our testing indicates that GROUNDING.md is authoritative largely due to its explicit language for HC (Appendix A.6.7, A.6.9, and A.6.10), but if such language is contained elsewhere in context it may cause conflicts. This testing also shows boundary cases in which compliance degrades under explicit override instructions or weakened normative language, underscoring that GROUNDING.md provides an auditable anchor and can substantially increase compliance under the tested conditions, but does not by itself guarantee correctness under all context conditions. Overall, formal adoption of this document and its utilization in agent scaffolds should alleviate such issues.

\section{Limitations}
There are limitations to using GROUNDING.md in its draft form. All tests in the present study were conducted in fresh, single-turn sessions, so durability over longer sessions with substantial intervening context remains untested. In addition, the present tests evaluate refusal under genuine violations, but do not measure false-positive refusals on valid requests or systematic auditing of pre-existing code with planted violations. The current study also uses a single proof-of-principle model configuration rather than a broader evaluation across default frontier-model deployments and additional local or open-weight models. Accordingly, this Letter should be interpreted as a proof of principle for a field-scoped epistemic grounding mechanism that encodes hard validity constraints, while broader benchmark-style evaluation of realistic tasks, longer sessions, false positives, and code-auditing behavior remains future work.

Beyond these empirical limits, the design of the GROUNDING.md framework has additional scope constraints. It has been described here as domain specific, but there may be multiple grounding files needed depending on the software development. For instance, though security (PII leakage) and project management are important they are intentionally excluded from the draft grounding document to keep the focus on proteomic functional correctness. We have also proposed (within the proteomics\_GROUNDING.md available at \url{https://github.com/OmicsGrounding/proteomics-grounding}) that subdomain extensions be maintained as separate documents but reference the primary domain document (i.e., metaproteomics\_GROUNDING.md would reference proteomics\_GROUNDING.md). Though more context files can provide the proposed domain specific HCs and CPs, in general there is a concern about the value of context files \cite{arxiv_2602_11988}. In our testing, providing HCs in GROUNDING.md accomplishes the field-scoped goal not possible without context engineering. Still, future work could optimize GROUNDING.md on the fly according to application or category (e.g., proteomics\_fdr\_GROUNDING.md).

A further limitation is practical rather than technical: authoring field-scoped grounding documents through expert consensus is itself time- and labor-intensive, and the community bodies best positioned to do this work move deliberately. While human- and community-governed authorship remains the eventual goal, and indeed the source of a GROUNDING.md's authority, waiting for fully consensus-vetted documents may slow adoption. As an interim measure, we propose that the substantial body of guidelines, best-practice papers, and reporting standards these organizations have already produced can serve as source material for LLMs to generate draft grounding files for other domains. Such drafts, clearly marked as provisional and posted publicly (e.g., on GitHub) for community review and refinement, could provide a usable starting point while human-led governance catches up, lowering the barrier to entry without bypassing the eventual need for expert validation.

It is also important to note other efforts to provide similar functionality as the proposed grounding document. For instance, Spec Kit provides a project-scoped specific constitution for agents to follow \cite{spec_kit}, while a recently added feature in Claude Code are \enquote{rules} being nested under project folders. In contrast, GROUNDING.md is field-scoped, epistemically founded, and community maintained is unique and accomplishes something new and different. Finally, using GROUNDING.md will vary by agent scaffold, but by using similar test prompts that violate HCs (Appendix A.3), we have demonstrated one way to demonstrate GROUNDING.md is being used as intended. This in turn can be used in future testing to demonstrate how consistent GROUNDING.md is followed or ignored since we cannot predict how model or agent scaffold updates will affect behavior.

\section{Summary and Future}
The idea of vibe coding and handing over the reins of software development to agentic AI is a divisive topic. It is our impression that this is already far more commonplace than generally known. For example, AI is reportedly now writing 90 \% of the lines of code at Anthropic \cite{dwarkesh_amodei} and it is the author’s knowledge that resources in the field of proteomics are being underpinned by software generated by agentic workflows. Moreover, non-experts making bespoke software generates a fair amount of community unease. But this democratization also presents an unexpected opportunity that if the research software developer uses an epistemic grounding document, then the process will adhere to domain specific guidelines, best practices, and documentation and reporting standards. Also, providing an aggregated set of directions that are straight forward for an agent to follow helps the research software developers confine the trajectory of the project while increasing the likelihood of a high-quality product. This is especially important for incorporating requirements for intermediate tests during development using public and in silico data, as well as providing conventions for best practices like statistical process control in ongoing quality control monitoring. For future hard constraints of software generated with GROUNDING.md, we encourage collaborative projects such as ProteoBench that specify how to benchmark and compare proteomics analysis workflows \cite{proteobench_2025}. We believe such a GROUNDING.md document will empower AI practitioners without proteomics expertise, novice proteomics researchers, or even experienced proteomics researchers taking advantage of the power of agentic AI coding to improve or customize existing tools.

The draft proteomic epistemic grounding document concept described herein is an attempt to capture several independent efforts from HUPO-PSI (Human Proteome Organization Proteomics Standards Initiative), journals, bioinformatics [e.g., DOME (Data, Optimization, Model, Evaluation) for machine learning \cite{jpr_2021_dome}] and research software community (Findable, Accessible, Interoperable, and Reusable for Research Software; FAIR4RS) recommendations and guidelines \cite{scientific_data_2022_fair}. We expect these groups could contribute to the development and maintenance of proteomic epistemic grounding documents, providing much more expertise than these authors can. Existing community outputs could be converted to a grounding document that is human-readable but agent-consumed containing HCs and CPs. Also, this document type could be used for other domains like biostatistical and systems analyses, or any prescriptive output, such as journal articles. It is also possible grounding documents could be generated specific to institutions and research groups, though this could drive the fragmentation that epistemic grounding is intended to combat.

By encoding domain community consensus on validity into GROUNDING.md, agentic workflows can adhere to field-scoped epistemic standards, preventing agents from optimizing for immediate goals while violating invariant constraints, even when guided by non-experts. Overall, the coming future of a ubiquitous agentic reality presents new challenges. An epistemic grounding document can take advantage of new opportunities for quality control, harmonization, and reproducibility, making the final products better for creators, reviewers, and end-users. As frontier models move toward the \enquote{country of geniuses} scale, the bottleneck for scientific agentic AI will shift from code generation to validity assurance, making field-scoped epistemic grounding documents like GROUNDING.md not just useful, but indispensable for trustworthy, reproducible science.

\section*{Disclaimer}
The text herein was not written by an LLM, though the example epistemic GROUNDING.md document itself was written with AI-assistance. An early draft was submitted to q.e.d. at https://www.qedscience.com/ to help identify areas for improvement. Identification of certain commercial equipment, instruments, software, or materials does not imply recommendation or endorsement by the National Institute of Standards and Technology or author-affiliated organizations, nor does it imply that the products identified are necessarily the best available for the purpose.

\section*{Conflicts of Interest}
MP, JMR, and BAN declare no competing financial interests.

\section*{Acknowledgments}
We want to thank Siegfried Gessulat for his invaluable conversations that prompted the initial idea for this paper.

\bibliographystyle{unsrt}
\bibliography{references}

@article{meyer2026,
  title={Vibe Coding Omics Data Analysis Applications},
  author={Meyer, Jesse G},
  journal={Journal of proteome research},
  volume={25},
  number={2},
  pages={1191--1197},
  year={2026}
}

@misc{amodei_essay,
  author = {Dario Amodei},
  title = {The Adolescence of AI},
  howpublished = {\url{https://www.darioamodei.com/essay/the-adolescence-of-technology}},
  year = {2024}
}

@misc{agents_md,
  title = {AGENTS.md},
  howpublished = {\url{https://agents.md/}}
}

@misc{skill_md,
  title = {SKILL.md},
  howpublished = {\url{https://skill.md/}}
}

@article{hupo_psi_2022,
  title={Proteomics standards initiative at twenty years: current activities and future work},
  author={Deutsch, Eric W and Vizca{\'\i}no, Juan Antonio and Jones, Andrew R and Binz, Pierre-Alain and Lam, Henry and Klein, Joshua and Bittremieux, Wout and Perez-Riverol, Yasset and Tabb, David L and Walzer, Mathias and others},
  journal={Journal of Proteome Research},
  volume={22},
  number={2},
  pages={287--301},
  year={2023},
  publisher={ACS Publications}
}

@article{arxiv_2212_08073,
  title={Constitutional ai: Harmlessness from ai feedback},
  author={Bai, Yuntao and Kadavath, Saurav and Kundu, Sandipan and Askell, Amanda and Kernion, Jackson and Jones, Andy and Chen, Anna and Goldie, Anna and Mirhoseini, Azalia and McKinnon, Cameron and others},
  journal={arXiv preprint arXiv:2212.08073},
  year={2022}
}

@article{arxiv_1606_06565,
  title={Concrete problems in AI safety},
  author={Amodei, Dario and Olah, Chris and Steinhardt, Jacob and Christiano, Paul and Schulman, John and Man{\'e}, Dan},
  journal={arXiv preprint arXiv:1606.06565},
  year={2016}
}

@misc{zenodo_18786768,
  author = {EVERSE Research Software Quality Kit (RSQKit)},
  title = {Zenodo Repository},
  year = {2024},
  doi = {10.5281/zenodo.18786768}
}

@misc{openai_skills,
  title = {OpenAI Skills},
  howpublished = {\url{https://github.com/openai/skills/tree/main}}
}

@misc{claude_skills,
  title = {Anthropic Claude Agent Skills Overview},
  howpublished = {\url{https://platform.claude.com/docs/en/agents-and-tools/agent-skills/overview}}
}

@misc{daily_dose_claude,
  title = {Anatomy of the Claude folder},
  howpublished = {\url{https://blog.dailydoseofds.com/p/anatomy-of-the-claude-folder}}
}

@article{jpr_2019_guidelines,
  title={Mass spectrometry-based plasma proteomics: considerations from sample collection to achieving translational data},
  author={Ignjatovic, Vera and Geyer, Philipp E and Palaniappan, Krishnan K and Chaaban, Jessica E and Omenn, Gilbert S and Baker, Mark S and Deutsch, Eric W and Schwenk, Jochen M},
  journal={Journal of proteome research},
  volume={18},
  number={12},
  pages={4085--4097},
  year={2019},
  publisher={ACS Publications}
}

@article{reference_materials,
  title={Quality control in the mass spectrometry proteomics core: a practical primer},
  author={Neely, Benjamin A and Perez-Riverol, Yasset and Palmblad, Magnus},
  journal={Journal of biomolecular techniques: JBT},
  volume={35},
  number={3},
  pages={3fc1f5fe--42308a9a},
  year={2024}
}

@article{nature_comm_2021,
  title={A proteomics sample metadata representation for multiomics integration and big data analysis},
  author={Dai, Chengxin and F{\"u}llgrabe, Anja and Pfeuffer, Julianus and Solovyeva, Elizaveta M and Deng, Jingwen and Moreno, Pablo and Kamatchinathan, Selvakumar and Kundu, Deepti Jaiswal and George, Nancy and Fexova, Silvie and others},
  journal={Nature Communications},
  volume={12},
  number={1},
  pages={5854},
  year={2021},
  publisher={Nature Publishing Group UK London}
}

@article{bbapap_2013,
  title={Data standardization and sharing—the work of the HUPO-PSI},
  author={Orchard, Sandra},
  journal={Biochimica et Biophysica Acta (BBA)-Proteins and Proteomics},
  volume={1844},
  number={1},
  pages={82--87},
  year={2014},
  publisher={Elsevier}
}

@article{nature_biotech_2007,
  title={The minimum information about a proteomics experiment (MIAPE)},
  author={Taylor, Chris F and Paton, Norman W and Lilley, Kathryn S and Binz, Pierre-Alain and Julian Jr, Randall K and Jones, Andrew R and Zhu, Weimin and Apweiler, Rolf and Aebersold, Ruedi and Deutsch, Eric W and others},
  journal={Nature biotechnology},
  volume={25},
  number={8},
  pages={887--893},
  year={2007},
  publisher={Nature Publishing Group US New York}
}

@inproceedings{eriksson_asms_1999,
  author = {Eriksson, Jan and Feny{\"o}, David and Chait, Brian},
  title = {On the quality of protein identification by mass spectrometry},
  booktitle = {Poster WPH 261 at ASMS 1999},
  year = {1999}
}

@article{pmid_11921442,
  title={A model of random mass-matching and its use for automated significance testing in mass spectrometric proteome analysis},
  author={Eriksson, Jan and Feny{\"o}, David},
  journal={Proteomics},
  volume={2},
  number={3},
  pages={262--270},
  year={2002},
  publisher={Wiley Online Library}
}

@article{nature_methods_2025,
  title={Assessment of false discovery rate control in tandem mass spectrometry analysis using entrapment},
  author={Wen, Bo and Freestone, Jack and Riffle, Michael and MacCoss, Michael J and Noble, William S and Keich, Uri},
  journal={Nature Methods},
  volume={22},
  number={7},
  pages={1454--1463},
  year={2025},
  publisher={Nature Publishing Group US New York}
}

@article{arxiv_2602_11988,
  title={Evaluating AGENTS. md: Are Repository-Level Context Files Helpful for Coding Agents?},
  author={Gloaguen, Thibaud and M{\"u}ndler, Niels and M{\"u}ller, Mark and Raychev, Veselin and Vechev, Martin},
  journal={arXiv preprint arXiv:2602.11988},
  year={2026}
}

@misc{spec_kit,
  author = {Delimarsky, Den and Riem, Manfred},
  title = {Spec Kit},
  howpublished = {\url{https://github.com/github/spec-kit}}
}

@misc{dwarkesh_amodei,
  author = {Dwarkesh Patel},
  title = {Dario Amodei - Dwarkesh Podcast},
  year = {2026},
  howpublished = {\url{https://www.dwarkesh.com/p/dario-amodei-2}}
}

@article{proteobench_2025,
  title={ProteoBench: the community-curated platform for comparing proteomics data analysis workflows},
  author={Devreese, Robbe and Jachmann, Caroline and Van Puyvelde, Bart and Anagho-Mattanovich, Holda A and Wolski, Witold E and Webel, Henry and Anagho-Mattanovich, Matthias and Bittremieux, Wout and Chaoui, Karima and Chiva, Cristina and others},
  journal={bioRxiv},
  pages={2025--12},
  year={2025},
  publisher={Cold Spring Harbor Laboratory}
}

@article{jpr_2021_dome,
  title={Interpretation of the DOME recommendations for machine learning in proteomics and metabolomics},
  author={Palmblad, Magnus and Bocker, Sebastian and Degroeve, Sven and Kohlbacher, Oliver and Kall, Lukas and Noble, William Stafford and Wilhelm, Mathias},
  journal={Journal of proteome research},
  volume={21},
  number={4},
  pages={1204--1207},
  year={2022},
  publisher={ACS Publications}
}

@article{scientific_data_2022_fair,
  title={Introducing the FAIR Principles for research software},
  author={Barker, Michelle and Chue Hong, Neil P and Katz, Daniel S and Lamprecht, Anna-Lena and Martinez-Ortiz, Carlos and Psomopoulos, Fotis and Harrow, Jennifer and Castro, Leyla Jael and Gruenpeter, Morane and Martinez, Paula Andrea and others},
  journal={Scientific data},
  volume={9},
  number={1},
  pages={622},
  year={2022},
  publisher={Nature Publishing Group UK London}
}

\end{document}